\documentclass[pra,aps,twocolumn,showpacs,a4paper,10pt,footinbib]{revtex4-1}
\usepackage{amsmath,amsfonts,amssymb,graphics,graphicx,epsfig,color,times}
\usepackage[latin1]{inputenc}
\usepackage{pifont}

\graphicspath{{Bilder/}}

\DeclareMathSymbol{\rho}{\mathord}{letters}{"25}
\DeclareMathSymbol{\varrho}{\mathord}{letters}{"1A}

\usepackage[plainpages=false,pdfpagelabels,colorlinks=true,citecolor=blue]{hyperref}								

\usepackage{cancel,ifthen}
\usepackage[normalem]{ulem}

\newcommand{\Komment}[2][NoInPuT]{\ifthenelse{\equal{#1}{NoInPuT}}{}{{\color{blue}\sout{#1}}}{\color{red} #2}}

\begin{document}

\title{Multi-band and nonlinear hopping corrections to the 3D Bose-Fermi-Hubbard model}

\author{Alexander Mering}
\affiliation{Fachbereich Physik and research center OPTIMAS, Technische Universit\"at Kaiserslautern, D-67663 Kaiserslautern, Germany}

\author{Michael Fleischhauer}
\affiliation{Fachbereich Physik and research center OPTIMAS, Technische Universit\"at Kaiserslautern, D-67663 Kaiserslautern, Germany}

\begin{abstract}
Recent experiments revealed the importance of higher-band effects for the Mott insulator (MI) -- superfluid transition (SF) of ultracold bosonic atoms or mixtures 
of bosons and fermions in deep optical lattices [Best \emph{et al.}, PRL \textbf{102}, 030408 (2009); Will \emph{et al.}, Nature \textbf{465}, 197 (2010)]. 
In the present work, we derive an effective lowest-band Hamiltonian in 3D that generalizes the standard Bose-Fermi Hubbard model taking these effects  as 
well as  nonlinear corrections of the tunneling amplitudes mediated by
interspecies interactions into account. It is shown that a correct description of the lattice states in terms of the bare-lattice Wannier 
functions rather than approximations such as harmonic oscillator states is essential.
In contrast to self-consistent approaches based on effective Wannier functions our approach captures the observed reduction
of the superfluid phase for repulsive interspecies interactions. 
\end{abstract}


\date{\today}

\maketitle

\section{Introduction}
Ultracold atoms in optical lattices provide unique and highly controlable realizations of various many-body Hamiltonians 
\cite{Jaksch1998,Joerdens2008,Schneider2008,Duan2003,Kuklov2003,Barthel2009}. Theoretical descriptions of these systems in the case of deep 
lattice potentials usually employ lowest-band models only \cite{Jaksch1998,Albus2003}. However, it was found recently that for lattice bosons with 
strong interaction contributions to the Hamiltonian beyond the single-band approximation with nearest-neighbor hopping and local two-particle 
interactions need to be taken into account \cite{Will2010}. E.g., using the method
of \emph{quantum phase diffusion}, the value of the two-body interaction $U$ for bosons in a deep optical lattice was measured directly and found 
to deviate from the prediction of the tight-binding model derived in \cite{Jaksch1998}. These experiments also revealed the presence of additional 
local three- and four-body interactions not accounted for in the single-band Bose-Hubbard Hamiltonian. A perturbative derivation of these terms 
based on harmonic-oscillator approximations was given by Johnson \textit{et al.} \cite{Johnson2009}.

In the case of boson-fermion mixtures, the situation is more involved. The first experiments on mixtures 
with attractive interspecies interaction \cite{Guenther2006,Ospelkaus2006} displayed a decrease of the bosonic superfluidity in the presence of fermions. 
This initiated a controversial discussion about the nature of the effect. Explanations ranged from localization effects of bosons induced by fermions
 \cite{Ospelkaus2006,Mering2008} to heating due to the admixture \cite{Guenther2006,Cramer2008}. Numerical results also predicted the opposite 
behavior, i.e., the enhancement of bosonic superfluidity due to fermions \cite{Pollet2008} with a more detailed discussion in \cite{Varney2008}. 
The situation remained unclear until a systematic experimental study of the dependence of the shift in the bosonic SF -- MI transition on the boson-fermion 
interaction \cite{Best2009} and the subsequent observation of higher-order interactions in the mixture. This shows, that again higher-order band effects 
need to be taken into account. 

The influence of higher Bloch bands in the Bose-Fermi mixture can be described by two different approaches: In the first approach one assumes that 
the single-particle Wannier functions are altered due to the modification of the lattice potential  for one species by the interspecies interaction with the 
other \cite{Luehmann2008}, which is then calculated in a self-consistent manner. The agreement of these results to experimentally observed shifts of 
the SF-MI transition is very good for the case of attractive boson-fermion interaction (see \cite{Best2009}). 
The method fails however for repulsive interactions where experiments showed contrary to intuition again a reduction of superfluidity \cite{Best2009}. 
Besides this shortcoming, the self-consistent potential approach has a conceptual weakness as it can only be applied close to the Mott-insulating phase. 
The second approach to include higher bands is an elimination scheme leading to an effective single-band Hamiltonian similar to the pure bosonic case 
\cite{Tewari2009,Johnson2009,Lutchyn2009}. This approach, although technically more involved, is more satisfactory from a fundamental point of view. 
It did not result in quantitatively satisfactory predictions so far, however. We will show here that this is because (i) an important non-linear correction to
 the hopping mediated by the
inter-species interaction and present already in absence of higher-band corrections has been missed out and (ii) harmonic oscillator approximations to 
the Wannier functions which have been used before, lead to gross errors when considering higher band effects.

We here present an adiabatic elimination scheme for Bose-Fermi mixtures obtained independently from \cite{Johnson2009,Tewari2009,Lutchyn2009}, 
resulting in an effective first-band BFH-Hamiltonian
\cite{MeringDPG2009}. In contrast to \cite{Johnson2009} and \cite{Tewari2009,Lutchyn2009} we use
correct Wannier functions, which will be shown to be
essential. Furthermore we find that already within the lowest Bloch band the inter-species interaction leads to 
important nonlinear corrections to the tunneling matrix elements of bosons and fermions.
For a fixed number of fermions per site, the effective Hamiltonian is equivalent to 
the Bose-Hubbard model with renormalized parameters $U$ and $J$ for which expressions are given 
in a closed form. This allows for a direct study of the influence of the boson-fermion interactions 
on the bosonic superfluid to Mott-insulator transition within this level of approximation. It is shown that
nonlinear hopping together with higher-band corrections lead to a reduction of the bosonic superfluidity 
when adding fermions for both, attractive {\it and} repulsive inter-species interactions.

The outline of the present work is as follows. After deriving the general multi-band Hamiltonian
of interacting spin-polarized fermions and bosons in a deep lattice in the following section, we 
introduce the first important addition to the standard BFHM in section \ref{Nonlinearhoppingcorrection}, the
nonlinear hopping correction. Restricting to leading contributions, we derive an effective single-band Hamiltonian
by adiabatic elimination of the higher bands 
in section \ref{sec:MultibandDerivation}. Finally, using the resulting generalized BFHM, the effect of a varying boson-fermion interaction is studied in 
detail in section \ref{sec:Evaluation}.

\section{model}

In 3D, ultracold Bose-Fermi mixtures in an external potential are described by the continuous Hamiltonian \cite{Albus2003}
\begin{equation}
\begin{split}
  \hat{H}&=\int {\rm d}^3r\, 
\hat\Psi_b^\dagger(\mathbf r)\left[-\frac{\hbar^2}{2m_b}\Delta+V^b(\mathbf r) 
\right]\hat\Psi_b(\mathbf r)\\
  &\hspace{0.5cm}+\int{\rm d}^3r\, \hat\Psi_f^\dagger(\mathbf r)\left[-\frac{\hbar^2}{2m_f}
\Delta+V^f(\mathbf r) \right]\hat\Psi_f(\mathbf r)\\
  &\hspace{0.5cm}+\frac{g_{\rm bb}}{2} \int {\rm d}^3r\, 
\hat\Psi_b^\dagger(\mathbf r)\hat\Psi_b^\dagger(\mathbf r)\hat\Psi_b(\mathbf r)\hat\Psi_b(\mathbf r)\\
  &\hspace{0.5cm}+\frac{g_{\rm bf}}{2} \int {\rm d}^3r\, 
\hat\Psi_b^\dagger(\mathbf r)\hat\Psi_f^\dagger(\mathbf r)\hat\Psi_f(\mathbf r)\hat\Psi_b(\mathbf r),\label{eq:ContinuousHamiltonianBoseFermi}
\end{split}
\end{equation}
where the index b (f) at the field operators $\hat\Psi$ refers to bosonic (fermionic) quantities and $V^b(\mathbf r)$ [$V^f(\mathbf r)$] is the external potential consisting of possible trapping potentials as well as the optical lattice $V_{\rm lat}^b(\mathbf r) = \eta_b\ \sum\sin^2(k_\alpha r_\alpha)$ [$V_{\rm lat}^f(\mathbf r) = \eta_f\ \sum\sin^2(k_\alpha r_\alpha)$]. The intra- and interspecies interaction constants are defined as
\begin{equation}
 g_{\rm bb}=\frac{4\pi\hbar^2}{m_b} \ a_{\rm bb},\hspace{1cm}  g_{\rm bf}=\frac{4\pi\hbar^2}{m_R} \ a_{\rm bf},
\end{equation}
with $m_R=\frac{m_b m_f}{m_b+m_f}$ being the reduced mass and $a_{\rm bb/bf}$ the intra- and interspecies $s$-wave scattering length, respectively.

Whereas in the standard approach the field operators in \eqref{eq:ContinuousHamiltonianBoseFermi} are expanded in terms of Wannier functions for the first band only, we here use an expansion to all Bloch bands:
\begin{equation}
\begin{split}
   \hat\Psi_b(\mathbf r) =\sum_{\boldsymbol\nu} \sum_{\mathbf j} \hat b_{\boldsymbol\nu,\mathbf j}\, w^{b}_{\boldsymbol\nu}(\mathbf r - \mathbf j ),\\
   \hat\Psi_f(\mathbf r) =\sum_{\boldsymbol\nu} \sum_{\mathbf j} \hat f_{\boldsymbol\nu,\mathbf j}\, w^{f}_{\boldsymbol\nu}(\mathbf r-\mathbf j).
\end{split}
\end{equation}
The operator $\hat b_{\boldsymbol\nu,\mathbf j}$ [$\hat f_{\boldsymbol\nu,\mathbf j}$] denotes the annihilation of a boson (fermion) in the $\boldsymbol\nu$-th band at site $\mathbf j$ and $w^{b/f}_{\boldsymbol\nu}(\mathbf r - \mathbf j )$ is the corresponding Wannier function of the $\boldsymbol\nu$-th band located at site $\mathbf j$. The vector $\boldsymbol\nu=\{\nu_x,\nu_y,\nu_z\}$ denotes the band index. The Wannier functions factorize as
\begin{equation}
 w^{b/f}_{\boldsymbol\nu}(\mathbf r )=\widetilde w^{b/f}_{\nu_x}(x )\ \widetilde w^{b/f}_{\nu_y}(y )\ \widetilde w^{b/f}_{\nu_z}(z )
\end{equation}
whith the one-dimensional Wannier function $\widetilde w^{b/f}_{\beta}(x )$.

 Using the expansion of the field operator, the full multi-band Bose-Fermi-Hubbard Hamiltonian can be expressed as:
\begin{gather}
\begin{split}
  \hat H &=\sum_{\substack{\boldsymbol\nu,\boldsymbol\mu\\{\mathbf j}_1,{\mathbf j}_2}}\left\{  J^{{\mathbf j}_1\mathbf j_2}_{\boldsymbol\nu\boldsymbol\mu} \, \hat b_{\boldsymbol\nu,{\mathbf j}_1}^\dagger \hat b_{\boldsymbol\mu,{\mathbf j}_2}+{\widetilde{J}}^{{\mathbf j}_1\mathbf j_2}_{\boldsymbol\nu\boldsymbol\mu} \hat f_{\boldsymbol\nu,{\mathbf j}_1}^\dagger \hat 
f_{\boldsymbol\mu,{\mathbf j}_2}\right\}\\
&\hspace{0.75cm}+\frac{1}2 \sum_{\substack{\boldsymbol\nu,\boldsymbol\mu,\boldsymbol\rho,\boldsymbol\sigma\\{\mathbf j}_1\dots\mathbf j_4}} \left\{  U^{{\mathbf j}_1\dots\mathbf j_4}_{\boldsymbol\nu\boldsymbol\mu\boldsymbol\rho\boldsymbol\sigma}\,     \hat b_{\boldsymbol\nu,{\mathbf j}_1}^\dagger \hat b_{\boldsymbol\mu,{\mathbf j}_2}^\dagger \hat b_{\boldsymbol\rho,{\mathbf j}_3} \hat b_{\boldsymbol\sigma,{\mathbf j}_4}\right\}\\
&\hspace{0.75cm}+ \frac{1}2 \sum_{\substack{\boldsymbol\nu,\boldsymbol\mu,\boldsymbol\rho,\boldsymbol\sigma\\{\mathbf j}_1\dots\mathbf j_4}}  \left\{V^{{\mathbf j}_1\dots\mathbf j_4}_{\boldsymbol\nu,\boldsymbol\mu,\boldsymbol\rho,\boldsymbol\sigma}   \hat b_{\boldsymbol\nu,{\mathbf j}_1}^\dagger \hat b_{\boldsymbol\mu,{\mathbf j}_2} \hat f_{\boldsymbol\rho,{\mathbf j}_3}^\dagger \hat f_{\boldsymbol\sigma,{\mathbf j}_4}\right\} .
\end{split}
\intertext{The generalized hopping amplitudes (still containing local energy contributions)}
 \begin{split}
  J^{{\mathbf j}_1\mathbf j_2}_{\boldsymbol\nu\boldsymbol\mu} &=\int {\rm d}^3r\, \bar w^b_{\boldsymbol\nu}(\mathbf r- \mathbf j_1)\ \times\\
  &\hspace{0.75cm}\times \left[-\frac{\hbar^2}{2m_b} \Delta+V^b(\mathbf r)\right]w_{\boldsymbol\mu}^b(\mathbf r-\mathbf j_2 ),
 \end{split}\label{eq:GeneralizedHoppingAmplitude-b}\\
  \begin{split}
  {\widetilde J}^{{\mathbf j}_1\mathbf j_2}_{\boldsymbol\nu\boldsymbol\mu} &=\int {\rm d}^3r\, \bar w^f_{\boldsymbol\nu}(\mathbf r- {\mathbf j}_1)\ \times\\
  &\hspace{0.75cm}\times\left[-\frac{\hbar^2}{2m_f}\Delta+V^f(\mathbf r)\right]w_{\boldsymbol\mu}^f(\mathbf r-\mathbf j_2 ),
  \end{split}\label{eq:GeneralizedHoppingAmplitude-f}
  \intertext{and the generalized interaction amplitudes}
  \begin{split}
  U^{{\mathbf j}_1\mathbf j_2\mathbf j_3\mathbf j_4}_{\boldsymbol\nu\boldsymbol\mu\boldsymbol\rho\boldsymbol\sigma} &= g_{\rm bb}\int {\rm d}^3r\,  \bar w^b_{\boldsymbol\nu}(\mathbf r- {\mathbf j}_1) \ \times\\
  &\hspace{0.75cm}\times \bar w^b_{\boldsymbol\mu}(\mathbf r-\mathbf j_2 ) w^b_{\boldsymbol\rho}(\mathbf r- \mathbf j_3)  w^b_{\boldsymbol\sigma}(\mathbf r-\mathbf j_4 ),
  \end{split}\label{eq:GeneralizedInteractionAmplitude-b}\\
  \begin{split}
    V^{{\mathbf j}_1\mathbf j_2\mathbf j_3\mathbf j_4}_{\boldsymbol\nu\boldsymbol\mu\boldsymbol\rho\boldsymbol\sigma} &= g_{\rm bf}\int {\rm d}^3r\,  \bar w^b_{\boldsymbol\nu}(\mathbf r- {\mathbf j}_1)  \ \times\\
  &\hspace{0.75cm}\times w^b_{\boldsymbol\mu}(\mathbf r-\mathbf j_2 ) \bar w^f_{\boldsymbol\rho}(\mathbf r-\mathbf j_3)  w^f_{\boldsymbol\sigma}(\mathbf r-\mathbf j_4 ),
  \end{split}\label{eq:GeneralizedInteractionAmplitude-f}
\end{gather}
are defined as usual. In the following we restrict our model in such a way, that only the most relevant terms are kept. 
Note that many of the matrix elements vanish because of the symmetry of the Wannier functions \cite{Kohn1959}.
 Unless stated otherwise we restrict ourselves to local 
 contributions in interaction terms, i.e. $\mathbf j_1=\dots=\mathbf j_4$ in $U^{\mathbf j_1,\dots ,\mathbf j_4}_{\boldsymbol\nu\boldsymbol\mu\boldsymbol\rho\boldsymbol\sigma} $ and
$V^{\mathbf j_1,\dots,\mathbf j_4}_{\boldsymbol\nu\boldsymbol\mu\boldsymbol\rho\boldsymbol\sigma} $ and in this case we drop the site indices.\\

With these restrictions, the general multi-band Hamiltonian can be cast in the following form
\begin{equation}
 \hat H = \hat H_{\mathbf1} + \sum_{\boldsymbol\nu\not=\mathbf1} \hat H_{\boldsymbol\nu}^{0}+\underset{\boldsymbol\nu,\boldsymbol\mu,\boldsymbol\rho,\boldsymbol\sigma}{{\sum}^\prime} 
\hat H_{\boldsymbol\nu\boldsymbol\mu\boldsymbol\rho\boldsymbol\sigma},\label{eq:fullMultiband}
\end{equation}
where the first term
\begin{equation}
\hat H_{\mathbf1} = \hat H_{\rm BFHM}+\hat H_{\rm nlin}
\end{equation}
describes the (pure) first-band ($\mathbf 1 = \{1,1,1\}$) dynamics consisting of the standard Bose-Fermi-Hubbard part $\hat H_{\rm BFHM}$ \cite{Albus2003} and nonlinear hopping corrections $\hat H_{\rm nlin}$ which will be discussed in the next section. 
The second term $\hat H_{\boldsymbol\nu}^{0}$ incorporates the (free) dynamics within the $\boldsymbol\nu$-th band and 
$\hat H_{\boldsymbol\nu\boldsymbol\mu\boldsymbol\rho\boldsymbol\sigma}$ describes the coupling between arbitrary bands $\boldsymbol\nu,\boldsymbol\mu,\boldsymbol\rho,\boldsymbol\sigma$. The prime in the sum indicates that at least one multi-index has to be different from the others. This general form of the full Hamiltonian serves as the starting point of our study.\\


\section{Nonlinear hopping correction}\label{Nonlinearhoppingcorrection}


Even when virtual transitions to higher bands are disregarded there are important corrections to the
standard BFHM if the boson-fermion interaction $V$ becomes large. The interspecies interaction term
in \eqref{eq:ContinuousHamiltonianBoseFermi} gives rize to a correction to the bosonic (and fermionic)
tunneling amplitude proportional to the occupation number of the corresponding complementary species.
These contributions, in the following termed as nonlinear hopping contributions, have been considered before \cite{Mazzarella2006,Amico2010}, but have been missed out in earlier discussions of corrections to the BFHM \cite{Tewari2009,Lutchyn2009}.

To establish notation let us recall first the usual single-band BFHM
\begin{equation}
\begin{split}
\hat H &= -J\  \sum_{\langle\mathbf i\mathbf j\rangle} \hat b_{\mathbf i}^\dagger \hat b_{\mathbf j} +\frac{U}2 \sum_{\mathbf j}   \hat n_{\mathbf j} 
\left(\hat n_{\mathbf j}-1\right)\\
   &\hspace{0.5cm}-{\widetilde J}\ \sum_{\langle\mathbf i\mathbf j\rangle}\hat f_{\mathbf i}^\dagger \hat f_{\mathbf j}+\frac{V}2\sum_{\mathbf j}  \hat n_{\mathbf j} \hat m_{\mathbf j}.
\end{split}
 \end{equation}
The amplitudes are determined by 
\begin{equation}
 U \equiv U^{\mathbf j\mathbf j\mathbf j\mathbf j}_{\mathbf1\mathbf1\mathbf1\mathbf1},\quad V \equiv V^{\mathbf j\mathbf j\mathbf j\mathbf j}_{\mathbf1\mathbf1\mathbf1\mathbf1},\quad J \equiv -J^{\mathbf j+\mathbf{\hat e},\mathbf j}_{\mathbf1,\mathbf1},\quad {\widetilde J}\equiv -{\widetilde J}^{\mathbf j+\mathbf{\hat e},\mathbf j}_{\mathbf1,\mathbf1}
\nonumber
\end{equation}
with $\mathbf{\hat e}$ being an unit vector in one of the three lattice directions. Due to the isotropic setup, the choice of the direction is irelevant.
From eq. \eqref{eq:GeneralizedInteractionAmplitude-b} and \eqref{eq:GeneralizedInteractionAmplitude-f}
two types of nonlinear hopping corrections arise: From the 
boson-boson interaction we obtain
 \begin{equation}
J^b_{\rm nl}  \sum_{\langle\mathbf i\mathbf j\rangle}\ \hat b_{\mathbf i}^\dagger \left( \hat n_{\mathbf i}+\hat n_{\mathbf j}\right) \hat b_{\mathbf j},\label{eq:NonlinearHoppingBosonic}
 \end{equation}
 whereas the boson-fermion interaction leads to both, bosonic and fermionic hopping corrections:
  \begin{equation}
  \begin{split}
  & { J}_{\rm nl}^f  \sum_{\langle\mathbf i\mathbf j\rangle}\ \hat b_{\mathbf i}^\dagger \left(\hat m_{\mathbf i}+\hat m_{\mathbf j}\right) \hat b_{\mathbf j}\\
   &\hspace{1cm}+{\widetilde J}_{\rm nl}  \sum_{\langle\mathbf i\mathbf j\rangle}\  \hat f_{\mathbf i}^\dagger \left( \hat n_{\mathbf i}+\hat n_{\mathbf j}\right) \hat f_{\mathbf j}.
  \end{split}\label{eq:NonlinearHoppingFermionic}
 \end{equation}
The corresponding nonlinear hopping amplitudes read
\begin{equation}
 J^b_{nl}\equiv U^{\mathbf j+\mathbf{\hat e},\mathbf j,\mathbf j,\mathbf j}_{\mathbf1\mathbf1\mathbf1\mathbf1},\quad 
 { J}_{nl}^f \equiv V^{\mathbf j+\mathbf{\hat e},\mathbf j,\mathbf j,\mathbf j}_{\mathbf1\mathbf1\mathbf1\mathbf1},\quad  {\widetilde J}_{nl} \equiv V^{\mathbf j,\mathbf j,\mathbf j+\mathbf{\hat e},\mathbf j}_{\mathbf1\mathbf1\mathbf1\mathbf1}.
\nonumber
\end{equation}

Since we are interested in the influence of the fermions to the bosons we assume in the following the fermions to be homogenously distributed. This assumption also used in \cite{Best2009, Luehmann2008} proved to be valid in the trap center and gives a considerable simplification. This amounts to replacing the fermionic number-operators by the fermionic filling: $\hat m_{\mathbf j} \to m$. Furthermore, the bosonic density-operators in eqns. \eqref{eq:NonlinearHoppingBosonic} and \eqref{eq:NonlinearHoppingFermionic} are replaced by the filling of the Mott-lobe under consideration, $\hat n_{\mathbf j} \to n$, for simplicity. 

 Alltogether, this allows us to write a  
Hamiltonian including corrections from the nonlinear hopping contributions. Defining the effective bosonic hopping amplitude as
 \begin{equation}
 J[n,m]\equiv J-2n\  {J_{nl}^b} -m {J}_{nl}^f,\label{eq:NormalizeHopping}
\end{equation}
the system is recast in the form of a pure BHM with density dependend hopping:
 \begin{equation}
\hat H_{\rm eff} = -J[n,m]\sum_{\langle\mathbf i\mathbf j\rangle}\   \hat a_{\mathbf i}^\dagger \hat a_{\mathbf j}+  \frac{U}2 \sum_{\mathbf j}\hat n_{\mathbf j} \left(\hat n_{\mathbf j}-1\right).
\end{equation}
Analyzing the resulting predictions for the MI-SF transition as a function of the filling and the interspecies interaction (see figure \ref{fig:TransitionShiftScatteringLengthBands}) one recognizes a substantial reduction of bosonic superfluidity for increasing interaction on the attractive side and a corresponding enhancement on the repulsive side, showing the importance of
nonlinear hopping terms for the precise determination of the MI--SF transition. Compared to the experimental results \cite{Best2009}, two main points arise. First,  although pointing into the right direction for attractive interactions, the overall shift is too small compared to the experimental observation. Second, for repulsive interactions, the transition is shifted to larger lattice depths, in contrast to the experimental findings.


\section{Effective single-band Hamiltonian}\label{sec:MultibandDerivation}

In the following we derive an effective single-band Hamiltonian that takes into account the coupling to higher bands.
The derivation is structured in the following way: We use an adiabatic elimination scheme presented in appendix \ref{sec:AdiabaticElimination} which reduces the main task to the calculation of the the second order cumulant $\langle\langle\mathcal T\  H_I(\tau+T)H_I(\tau)\rangle\rangle$ in the interaction picture, where the average is taken over the higher bands. The full interaction Hamiltonian $\hat H_I=\underset{\boldsymbol\nu,\boldsymbol\mu,\boldsymbol\rho,\boldsymbol\sigma}{{\sum}^\prime} \hat H_{\boldsymbol\nu\boldsymbol\mu\boldsymbol\rho\boldsymbol\sigma}$ is then reduced according to the relevant contributions of the cumulant. Finally, a reduction of the effective bosonic scattering matrix \eqref{eq:SMatrixMultibandZwo} gives the full effective single-band Bose-Fermi-Hubbard model.\\

When calculating the cumulant $\langle\langle\mathcal T\  H_I(\tau+T)H_I(\tau)\rangle\rangle$ in \eqref{eq:SMatrixMultibandZwo}, the interaction Hamiltonian of the full multi-band Bose-Fermi-Hubbard model can be reduced considerable. Keeping only terms that lead to non-zero contributions in lowest order, it is easy to see that only those terms in $\hat H_I$ matter, where particles are transfered to higher bands by $\hat H_I(\tau)$ and down again by $\hat H_I(\tau+T)$.  In the following we restrict ourselves to precisely those contributions and furthermore treat only local contributions since these are dominant. Three relevant processes are found:
\begin{enumerate}
 \item Single particle transitions to a certain band $\boldsymbol\nu$
 \begin{center}
  $\{\mathbf1,\mathbf1,\mathbf1,\mathbf1\}\hspace{0.5cm}\leftrightarrow\hspace{0.5cm}\{\boldsymbol\nu,\mathbf1,\mathbf1,\mathbf1$\}
 \end{center}
 These contributions can be understood as density-mediated band transitions, where the matrix elements $U^{}_{\boldsymbol\nu\mathbf1\mathbf1\mathbf1}$, $V^{}_{\boldsymbol\nu\mathbf1\mathbf1\mathbf1}$, $V^{}_{\mathbf1\boldsymbol\nu\mathbf1\mathbf1}$ are only non-zero for odd bands $\boldsymbol\nu$. \footnote{Odd means in the 3D system that the number of odd elements in the multi-index $\boldsymbol\nu$ has to be odd!} Note that from now on, the upper site-indices are omitted if they are all the same.

\item Double-transition to the same band $\boldsymbol\nu$
 \begin{center}
 $\{\mathbf1,\mathbf1,\mathbf1,\mathbf1\}\hspace{0.5cm}\leftrightarrow\hspace{0.5cm}\{\boldsymbol\nu,\boldsymbol\nu,\mathbf1,\mathbf1\}$
 \end{center}

In this situations, two particles undergo a transition to the same band and all bands are incorporated. The matrix elements are $U_{\boldsymbol\nu\boldsymbol\nu\mathbf1\mathbf1}$ and $V_{\boldsymbol\nu\boldsymbol\nu\mathbf1\mathbf1}$.

\item Double-transition to different bands $\boldsymbol\nu,\boldsymbol\mu$
 \begin{center}
 $ \{\mathbf1,\mathbf1,\mathbf1,\mathbf1\}\hspace{0.5cm}\leftrightarrow\hspace{0.5cm}\{\boldsymbol\nu,\boldsymbol\mu,\mathbf1,\mathbf1\}$
 \end{center}

In this combined process, the two different bands have to be both either even or odd with matrix elements $U_{\boldsymbol\nu\boldsymbol\mu\mathbf1\mathbf1}$ and $V_{\boldsymbol\nu\boldsymbol\mu\mathbf1\mathbf1}$

\end{enumerate}

The remaining important contributions to the full multi-band BFHM result from the kinetic energy of the particles. Restricting to the usual nearest neigbour
hoppings within a given Bloch band ($\boldsymbol\nu=\boldsymbol\mu$ and $|\mathbf j_1-\mathbf j_2|=1$) and the energy of the particles within a band ($\boldsymbol\nu=\boldsymbol\mu$ and $\mathbf j_1=\mathbf j_2$), these are  

  \begin{figure}[t]
  \centering
\epsfig{file=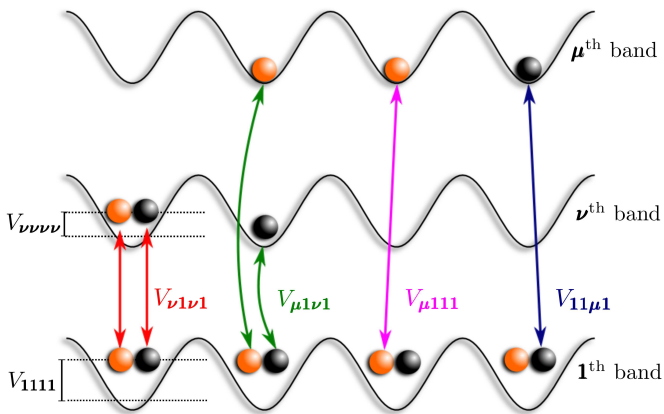,width=1\columnwidth}
   \caption{(Color online) Matrix elements for the coupling of the higher Bloch bands to the first band via the 
generalized interaction \eqref{eq:GeneralizedInteractionAmplitude-f}. Bosonic contributions from \eqref{eq:GeneralizedInteractionAmplitude-b} are equivalent. Bosons are shown as orange circles and fermions in black. 
{\color{red} $V_{\boldsymbol\nu\mathbf1\boldsymbol\nu\mathbf1}$} describes the transition of a boson and a fermion from the first 
(higher) to the higher (first) band; {\color[rgb]{0, 0.5, 0} $V_{\boldsymbol\mu\mathbf1\boldsymbol\nu\mathbf1}$} gives two particles 
(boson and fermion) which perform a transition to bands $\boldsymbol\nu$ and $\boldsymbol\mu$. {\color{magenta} $V_{\boldsymbol\mu\mathbf1\mathbf1\mathbf1}$} derscribes a fermion-mediated single particle transition of a boson, where 
{\color{blue} $V_{\mathbf1\mathbf1\boldsymbol\mu\mathbf1}$} is a boson-mediated transition of a fermion.}
   \label{fig:ProcessesGeneralizedInteraction}
    \end{figure}
\begin{enumerate}
\setcounter{enumi}{3}
 \item the band energies $\Delta^b_{\boldsymbol\nu}$ and $\Delta^f_{\boldsymbol\nu}$ 
\item the intraband nearest-neighbor hopping for bosons $J_{\boldsymbol\nu}$ and correspondingly for the fermions ${\widetilde J}_{\boldsymbol\nu}$.
\end{enumerate}
Hopping between sites with $|\mathbf j_1-\mathbf j_2|\not=1$ is omitted since it is unimportant. In appendix \ref{sec:AppendixCouplings}, the different contributions to the Hamiltonian as well as the hoppings and band energies are defined in detail. Figure \ref{fig:ProcessesGeneralizedInteraction} gives a sketch of the different contributions taken into
account. Shown are only processes involving fermions.\\

From the effective bosonic scattering matrix in \eqref{eq:SMatrixMultibandZwo}, the effective single-band BFHM is derived by applying a Markov approximation \cite{Carmichael1993}. This amounts to replacing first-band operators at time $\tau+T$ by the corresponding operators at time $\tau$ which is valid since the timescale of the higher-band dynamics is much shorter than in the first band because of the larger hopping amplitude \cite{Isacsson2005}. The resulting Hamiltonian is lengthy and shows the full form is given in appendix \ref{sec:AppendixFullHamiltonian}.

The effective Hamiltonian \eqref{eq:BthBandEffectieHamiltonian} contains non-local interaction and long-range tunneling terms.
These result from virtual transitions into higher bands and subsequent tunneling processes in these bands. 
As these terms rapidly decrease with increasing distance $|\mathbf d|$ between the involved lattice sites, it is sufficient to take 
into account only the leading order contributions, i.e. only local interaction terms ($|\mathbf d|=0$) and only nearest neigbour
hopping $(|\mathbf d|=\pm 1)$.  This leads to the following extensions compared to the standard single-band BFHM:
 \begin{widetext}
\begin{align}
 \hat H^{\rm eff} &= \sum_{\mathbf j} \Bigl\{\frac{U_3}6\  \hat n_{\mathbf j} \left(\hat n_{\mathbf j}-1\right)\left(\hat n_{\mathbf j}-2\right) + \frac{V_3}{2}\ 
\hat m_{\mathbf j}\hat n_{\mathbf j} 
\left(\hat n_{\mathbf j}-1\right)+  \frac{U_2}2\hat n_{\mathbf j} \left(\hat n_{\mathbf j}-1\right) +\frac{V_2}{2} \hat n_{\mathbf j}\hat m_{\mathbf j}\Bigr\} 
\label{eq:EffectivFirstbandHamiltonain}\\
 &\hspace{0.5cm}+ \sum_{\mathbf j} \Delta_{\mathbf1}^b \hat n_{\mathbf j} + \sum_{\mathbf j} \Delta_{\mathbf 1}^b \hat m_{\mathbf j}  - \sum_{\langle\mathbf i\mathbf j\rangle} \hat b_{\mathbf i}^\dagger\  J[\hat n_{\mathbf i},\hat n_{\mathbf j},\hat m_{\mathbf i},\hat m_{\mathbf j}]\   \hat b_{j}-\sum_{\langle\mathbf i\mathbf j\rangle} 
 \hat f_{\mathbf i}^\dagger\  {\widetilde J}[\hat n_{\mathbf i},\hat n_{\mathbf j}]\   \hat f_{\mathbf j}+ \sum_{\langle\mathbf i\mathbf j\rangle}\left\{ J^{(2)}  \left(\hat b_{\mathbf i}^\dagger \right)^2 \hat b_{\mathbf j}^2 + {\widetilde J}^{(2)}\   \hat b_{\mathbf i}^\dagger 
\hat f_{\mathbf i}^\dagger\hat f_{\mathbf j} \hat b_{\mathbf j}\right\}.\notag
 \end{align}
\end{widetext}

Here some new terms arize, for instance correlated two-particle tunneling $J^{(2)}$ and ${\widetilde J}^{(2)}$. Most prominent is the appearance of the three-body interactions $U_3$ and $V_3$.The bosonic has recently been measured by means of quantum phase diffusion \cite{Will2010}. It should be noted that in the experiments in \cite{Will2010} also higher order nonlinear interactions were detected. Since our approach is only second order in the interaction-induced intra-band coupling, these terms cannot be reproduced however. Beside the new terms, the higher bands lead to a renormalization of the usual single-band BFHM parameters. Whereas the local two-body interaction amplitudes $U_2$ and $V_2$ only depend on the band structure, the hopping amplitudes are altered, leading to density mediated hopping processes. For the bosonic ones, the hopping now is of the form
\begin{multline}
 J[\hat n_{\mathbf i},\hat n_{\mathbf j},\hat m_{\mathbf i},\hat m_{\mathbf j}]= J-J_{nl}^b\left(\hat n_{\mathbf j}+\hat n_{\mathbf i}\right)\\
 -\frac{ J_{nl}^f }{2}\ \left(\hat m_{\mathbf j}+\hat m_{\mathbf i}\right)+\alpha\ \hat n_{\mathbf i}\hat n_{\mathbf j}\\
 +\beta\ \hat m_{\mathbf i}\hat n_{\mathbf j}+\gamma\ \hat n_{\mathbf i}\hat m_{\mathbf j}+\delta\ \hat m_{\mathbf i}\hat m_{\mathbf j}
\end{multline}
and the density dependence is directly seen. For all parameters occuring in \eqref{eq:EffectivFirstbandHamiltonain}, full expressions can be found in appendix \ref{sec:AppendixDefinition}.

\section{Influence of fermions on the bosonic MI--SF transition}\label{sec:Evaluation}

In order to discuss the phase transition of the bosonic subsystem, we make further approximations. Coming from the Mott insulator
side of the phase transition, the local number of bosons is approximately given by the integer average filling, i.e., $\langle \hat n_{\mathbf j} \rangle \approx n$. For the fermionic species, we also replace the number operator by the average fermion number  $\hat m_{\mathbf j} \to m=1$, assuming a homogeneous filling of fermions in the lattice. Having an experimental realization with cold atoms in mind, this is a valid assumption in the center of the harmonic trap at least for attractive inter-species interactions. It should be valid however also for slight inter-species repulsion. This assumption is also supported by the results of \cite{Best2009}, where the actual fermionic density did not influence the transition from a Mott-insulator to a superfluid (for medium and large filling). It also agrees with the result in \cite{Luehmann2008} which is based on this assumption, and which shows a good agreement to the experimental results. All further contributions in the Hamiltonian 
such as the bosonic three-particle interaction  and two-particle hoppings are neglected in the following. With these approximations, the renormalized Bose-Hubbard Hamiltonian for the $n$-th Mott lobe with mean fermionic filling $m$ can be written
as
\begin{align}
 \hat H^{\rm eff} &= -J[n,m]\sum_{\langle\mathbf i\mathbf j\rangle}\ \hat b_{\mathbf i}^\dagger \hat b_{\mathbf j}+  \frac{U[m]}2 \sum_{\mathbf j}\hat n_{\mathbf j} \left(\hat n_{\mathbf j}-1\right)
\end{align}
with
\begin{align}
J[n,m]&=J-2 n \ J_{nl}^b - m\ J_{nl}^f \label{eq:correctedHopping}\\
&\hspace{1cm}-\sum_{\boldsymbol\nu\not=\mathbf1}\mathcal I^{\mathbf{\mathbf{\hat e}}}_{b,\boldsymbol\nu}\left(U_{\boldsymbol\nu\mathbf1\mathbf1\mathbf1}^{}\  n 
+ V_{\boldsymbol\nu\mathbf1\mathbf1\mathbf1}^{}\frac{m}{2} \right)^2\notag\\
U[m]&=U_2+m\ V_3\label{eq:correctedInteraction} \end{align}
The final form of the bosonic Hamiltonian will now be used to discuss the influence of the boson-fermion interaction on the Mott-insulator to superfluid transition. Following the experimental procedure presented in \cite{Best2009}, we consider the shift of the bosonic transition
 as a function of the boson-fermion interaction determined by the scattering length $a_{BF}$, with a special emphasis on repulsive interaction where no theoretical prediction exists so far.

  \begin{figure}[t]
  \centering
   \epsfig{file=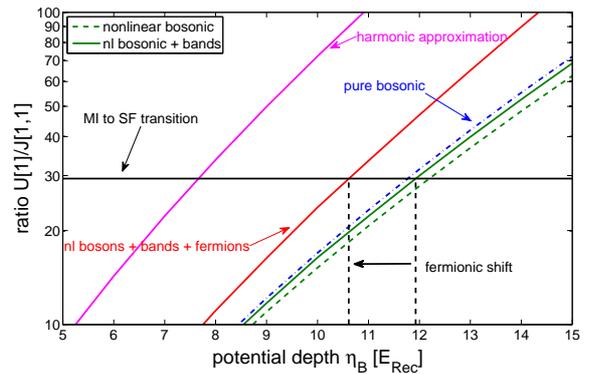,width=\columnwidth}
  \caption{(Color online) Ratio of effective interaction $U$ to effective tunneling rate $J$ for
unity fermion filling $m=1$ and Mott lobe with $n=1$ as function of
normalized lattice depth $\eta_b$ for the bosons, and for attractive boson-fermion interaction with a scattering length $a_{bf}=-400\ a_0$. The horizontal dotted line gives the critical
value for the MF -- SF transition point in the Bose-Hubbard model. Shown are the harmonic 
oscillator approximation together with different levels of corrections  as described in the main text based on exact Wannier functions. Assumed is a perfect match between fermionic and bosonic Wannier functions, $\eta_F\equiv\eta_B$.}
   \label{fig:DeterminationTransitionPointBands}
    \end{figure}

To determine the transition point, we calculate the bosonic hopping \eqref{eq:correctedHopping} and interaction amplitude \eqref{eq:correctedInteraction} using numerically determined Wannier functions. The knowlegde of the critical ratio $U/J$ of the MI to SF transition from analytic or numerical results \cite{Teichmann2009,dosSantos2009,Capogrosso-Sansone2007} allows for the precise localization of the transition point \cite{Greiner2002}. 
This method is displayed in figure \ref{fig:DeterminationTransitionPointBands} where the ratio of the effective interaction strength $U[1]$ and the effective tunneling rate $J[3,1]$ as per eq. \eqref{eq:correctedHopping} to \eqref{eq:correctedInteraction}  are plotted as a function of the normalized lattice depth $\eta_b$, which describes the amplitude of the periodic lattice potential of the
bosons $V^b_{\rm lat}$ in units of the recoil energy of the bosons $E_{\rm rec}^b = \hbar^2 \mathbf{k}^2/(2 m_b)$.  As indicated, unity fermion filling $m=1$ is assumed and the bosonic Mott lobe with $n=3$ is considered.
The horizontal dotted line gives the critical value for the MI -- SF transition \cite{Teichmann2009} and the crossing of this line with the different curves, which
illustrate the relative contribution of the various correction terms, determines the potential depth at which the phase transition occurs. The different levels of approximation shown in figure \ref{fig:DeterminationTransitionPointBands} are
\begin{enumerate}
 \item harmonic oscillator:\\
 plain BHM, harmonic oscillator approximation
 \item pure bosonic:\\
 plain BHM, proper Wannier functions
 \item nonlinear bosonic:\\
 BHM extended by nonlinear (bosonic) hopping correction
 \item nonlinear bosonic with higher bands:\\
 inclusion of all bands with $\nu_\alpha \le 25$; this gives the reference point for the shift of the transition
 \item nonlinear bosonic and fermionic with higher bands:\\
 inclusion of fermions; nonlinear hopping correction and higher bands ($\nu_\alpha \le 25$)
\end{enumerate} 
One clearly recognizes a substantial shift of the transition
point to lower potential depth in qualitative agreement with the experiment. 
It is also apparent that using harmonic oscillator approximations leads to a large error of the 
predicted transition point. This shows that the use
of the correct Wannier functions is crucial for obtaining reliable predictions.

\begin{figure}[t]
  \centering
    \epsfig{file=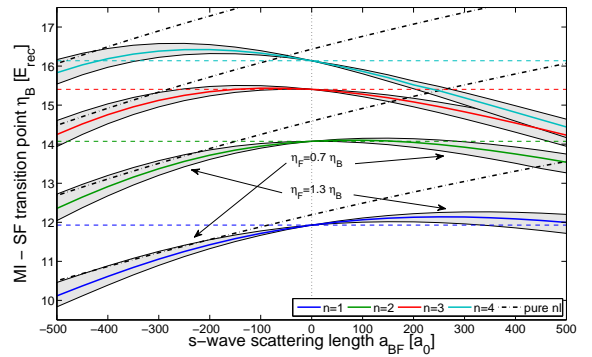,width=\columnwidth}
 \caption{(Color online) Shift of the bosonic Mott-insulator to superfluid transition as a function of the boson-fermion 
scattering length $a_{bf}$ for different Mott lobes (solid lines, $n=1\dots4$, from bottom to top) in one dimension. The gray-shaded region depicts the influence of a mismatch of the bosonic and fermionic lattice depth. The dot-dashed lines give the shifts of the transition solely from the nonlinear tunneling corrections. Dashed horizontal lines give the transition points for the pure bosonic system.}
   \label{fig:TransitionShiftScatteringLengthBands}
    \end{figure}

Figure \ref{fig:TransitionShiftScatteringLengthBands} shows the shift of the MI -- SF transition point 
for the first four lobes as a 
function of the boson-fermion scattering length $a_{bf}$. The solid lines include all corrections described earlier, where the amount of the shift is measured relative to the nonlinear bosonic case including higher bands, i.e., 
relative to the real bosonic transition point. Thus the figure corresponds to the shift of the transition point
when fermions with unity filling are added to the system. For each Mott lobe three curves are shown corresponding to different 
ratios of $\eta_f/\eta_b$ which illustrates the effect of different masses and/or different polarizabilities of the
bosonic and fermionic species as discussed in Appendix \ref{sec:AppendixLatticeEffects}. The dashed-dotted curves give the contributions of the (first band) nonlinear hopping corrections only (bosons and fermions). One recognizes that for increasingly attractive interactions between the species
there is an increasing shift of the transition point towards smaller potential depth, corresponding to a
reduction of bosonic superfluidity in the presence of fermions. Interestingly one recognizes that for
repulsive interspecies interactions, virtual transitions to higher Bloch bands tend to counteract the
effect of the fermion induced nonlinear tunneling. For larger values of $n$ there is again a shift
of the MI -- SF transition point towards smaller lattice depth, i.e. again a {\it reduction} of bosonic
superfluidity! The latter effect has both been observed in the experiments \cite{Best2009}, but has not been fully 
understood so far. In the calculations, the bands are summed up to a maximal multi-index $\boldsymbol\nu_{\rm max} = \{25,25,25\}$, including altogether 15625 bands. For this number of bands, a satisfactory convergence of the effective amplitudes $U$ and $J$ is found. Overall, our second order approach inlcuding the nonlinear corrections already provides an intuitive explanation for the behaviour of the system in the experiment. This especially holds for the repulsive case, where the agreement to the experimental results is on a quantitative level.

\section{Summary and outlook}

In the present paper we studied the influence of nonlinear tunneling processes and higher Bloch bands
on the dynamics of a mixture of bosons and fermions in a deep optical lattice in a full 3D setup. Taking into account virtual 
inter-band transitions in lowest non-vanishing order and contributions of the originally continuous
interaction to tunneling processes we derived an an effective lowest-band Hamiltonian
extending the standard Bose-Fermi Hubbard model. This Hamiltonian
contains interaction-mediated nonlinear corrections to the tunneling rates, renomalized two-body
interactions, and effective three-body interaction terms.  We showed that 
an accurate determination of the effective model parameter requires the use of the correct Wannier
functions of the corresponding single-particle model. As differences in the tails of the wavefunctions
are essential, the use of approximate harmonic oscillator wavefunctions can lead to large errors.
The effective model allows for a study of the effect of admixing
spin-polarized fermionic atoms to the bosonic superfluid to Mott-insulator transition when changing the
boson-fermion interaction strength. 
Our model recovers qualitatively all features observed in the experiment.
In particular we found that boson superfluidity is reduced both for attractive {\it and} repulsive
inter-species interactions. The latter has not been reproduced so far with other methods such as
the self-consistent potential approach. 

It should be noted that our model does not take into account heating effects and effects such as
phase separation due to the presence of an inhomogeneous trapping potential, which have recently been
shown  to significantly affect the MI-SF transition point already in the lowest band \cite{Cramer2010,Snoek2010}. We thus expect 
that a complete 
picture of the experimental observations will require a proper inclusion of higher-band effects and nonlinear tunneling as 
derived in the present paper, as well as effects from heating and a trapping potential. 
Finally it should be mentioned that our approach is limited to the second order in intra-band processes. 
In higher-order perturbation theory effective four-body, five-body, etc. interactions will arise, which play however
a less and less important role. Nevertheless, we expect that the higher orders should substantially improve the results, especially for 
repulsive interactions. 

\section*{Acknolwedgement}

The authors thank  S. Das Sarma, I. Bloch and E. Demler for useful discussions. The financial support by the
DFG through the SFB-TR49 is gratefully acknowledged. 


\begin{appendix}
\section*{Appendix}

\section{Adiabatic elimination scheme}\label{sec:AdiabaticElimination}

As long as the interaction energies $U$ and $V$ as well as the temperature are small 
compared to the band gap between lowest and first excited Bloch band, the population of 
higher bands can be neglected. However, as noted before, there are
virtual transitions to higher bands which need to be taken into account. 
In the following we employ an adiabatic elimination scheme of higher Bloch bands 
starting from the general multiband Hamiltonian \eqref{eq:fullMultiband}.  This scheme, which is also used in \cite{Mering2010} for the Bose-Fermi-Hubbard model in the ultrafast-fermion limit, is equivalent to degenerate perturbation theory \cite{Klein1974,Teichmann2009} and allows for a proper description of the reduced system. For this,
the Hamiltonian \eqref{eq:fullMultiband} is 
split up into a free and an interaction part $ \hat H = \hat H_{free} +\hat H_I$ with
\begin{align}
\hat H_{free} &=\hat  H_{\mathbf1} + \sum_{\boldsymbol\nu\not=\mathbf1}\hat H_{\boldsymbol\nu}^0,\\
 \hat H_I &= \underset{\boldsymbol\nu,\boldsymbol\mu,\boldsymbol\rho,\boldsymbol\sigma}{{\sum}^\prime}\hat H_{\boldsymbol\nu,\boldsymbol\mu,\boldsymbol\rho,\boldsymbol\sigma}.
\end{align}
Transforming to the interaction picture, the dynamics of the free part is incorporated by the time dependent interaction Hamiltonian $ H_{I}(\tau) = e^{-\frac{i}{\hbar} H_{free}\tau}\   H_I\  e^{\frac{i}{\hbar}  H_{free}\tau}$. Adiabatic elimination is carried out for the time evolution operator (scattering matrix) of the full system given by
\begin{equation}
 {\mathcal S} ={ \mathcal T}\exp\left\{-\frac i\hbar \int_{-\infty}^\infty {\rm d}\tau   \, \hat H_I(\tau)\right\}. 
\label{eq:SMatrixMultiband}
\end{equation}
We now trace out the higher-band degrees of freedom, assuming empty higher bands.  Using Kubo's cumulant expansion \cite{Kubo1962}
\begin{equation}
 \left \langle \exp \{ s X \} \right\rangle_{\rm X} = \exp \left\{\sum_{m=1}^\infty \frac{s^m}{m!} \langle\langle 
X^m\rangle\rangle \right\}\label{eq:CumulantExpansion}
\end{equation}
up to second order in the interband coupling, the effective scattering matrix for the lowest band reads
\begin{align}
{\mathcal S}_{\rm eff} &= \mathcal T \exp \Biggl\{ \label{eq:SMatrixMultibandZwo}\\
 &\hspace{-0.5cm}+\frac{1}{2} \left(- \frac{i}{ \hbar}\right)^2 {\int_{-\infty}^\infty {\rm d}\tau}
 \int_{-\infty}^\infty {\rm d}T\langle\langle\mathcal T\  H_I(\tau+T)H_I(\tau)\rangle\rangle \Biggr\}\notag.
\end{align}
The first order does not lead to any contributions because of the vacuum in the higher bands and due to the nature of the interband couplings. Obviously the effective bosonic Hamiltonian is connected to the second order cumulants of
operators in higher Bloch bands, $\langle\langle \hat A \ \hat B\rangle\rangle=\langle \hat A\ \hat B\rangle-\langle\hat A\rangle\langle\hat B\rangle$ \cite{Kubo1962}.

\section{Relevant band-coupling processes}\label{sec:AppendixCouplings}

As discussed in section \ref{sec:MultibandDerivation}, the different terms to the Hamiltonian are given by
\begin{enumerate}

 \item Single particle transitions to a certain band $\boldsymbol\nu$
 \begin{equation}
\begin{split}
 &\frac12\sum_{\mathbf j} \left[ U^{}_{\boldsymbol\nu\mathbf1\mathbf1\mathbf1} \  \hat b_{\mathbf1}^\dagger\hat b_{\mathbf1}  \hat b_{\boldsymbol\nu}^\dagger\hat b_{\mathbf1}+  U^{}_{\mathbf1\boldsymbol\nu\mathbf1\mathbf1} 
\ \hat b_{\mathbf1}^\dagger\hat b_{\mathbf1} \hat b_{\boldsymbol\nu}^\dagger\hat b_{\mathbf1} +\right.\\
 &\hspace{1cm} \left.+ U^{}_{\mathbf1\mathbf1\boldsymbol\nu\mathbf1}\  \hat b_{\mathbf1}^\dagger \hat b_{\boldsymbol\nu} \hat b_{\mathbf1}^\dagger \hat b_{\mathbf1}+ U^{}_{\mathbf1\mathbf1\mathbf1\boldsymbol\nu} 
 \ \hat b_{\mathbf1}^\dagger\hat b_{\boldsymbol\nu}  \hat b_{\mathbf1}^\dagger \hat b_{\mathbf1}+ \right.\\
&\hspace{1cm} \left. + V^{}_{\boldsymbol\nu\mathbf1\mathbf1\mathbf1} \  \hat b_{\boldsymbol\nu}^\dagger\hat b_{\mathbf1}  \hat f_{\mathbf1}^\dagger\hat f_{\mathbf1}+  V^{}_{\mathbf1\boldsymbol\nu\mathbf1\mathbf1} 
\ \hat b_{\mathbf1}^\dagger\hat b_{\boldsymbol\nu} \hat f_{\mathbf1}^\dagger\hat f_{\mathbf1} +\right.\\
 &\hspace{1cm} \left.+ V^{}_{\mathbf1\mathbf1\boldsymbol\nu\mathbf1}\  \hat b_{\mathbf1}^\dagger \hat b_{\mathbf1} \hat f_{\boldsymbol\nu}^\dagger \hat f_{\mathbf1}+ V^{}_{\mathbf1\mathbf1\mathbf1\boldsymbol\nu} 
 \ \hat b_{\mathbf1}^\dagger\hat b_{\mathbf1}  \hat f_{\mathbf1}^\dagger \hat f_{\boldsymbol\nu} \right].
\end{split}\label{eq:ContributionSingle}
\end{equation}
\item Double-transition to the same band $\boldsymbol\nu$
 \begin{equation}
\begin{split}
&\frac12\sum_{\mathbf j} \left[ U_{{\boldsymbol\nu} {\boldsymbol\nu}{\mathbf1}{\mathbf1}}^{}\ \hat b_{{\boldsymbol\nu}}^\dagger \hat b_{{\boldsymbol\nu}}^\dagger \hat b_{{\mathbf1}} \hat b_{{\mathbf1}}   + U_{{\mathbf1}{\mathbf1}{\boldsymbol\nu}{\boldsymbol\nu}}^{}\ \hat b_{{\mathbf1}}^\dagger \hat b_{{\mathbf1}}^\dagger \hat b_{{\boldsymbol\nu}} \hat b_{{\boldsymbol\nu}}+\right.\\ 
&\hspace{1 cm} \left. + V_{{\boldsymbol\nu} {\mathbf1}{\boldsymbol\nu} {\mathbf1}}^{}\ \hat b_{{\boldsymbol\nu}}^\dagger \hat b_{{\mathbf1}} \hat f_{{\boldsymbol\nu}}^\dagger \hat f_{{\mathbf1}}   + V_{{\mathbf1}{\boldsymbol\nu}{\mathbf1}{\boldsymbol\nu}}^{}\ \hat b_{{\mathbf1}}^\dagger \hat b_{{\boldsymbol\nu}} \hat f_{{\mathbf1}}^\dagger \hat f_{{\boldsymbol\nu}}\right].
\end{split}
\end{equation}

\item Double-transition to different bands ${\boldsymbol\nu},\boldsymbol\mu$
 \begin{equation}
\begin{split}
&\frac12\sum_{\mathbf j} \left[ U_{{\boldsymbol\nu} \boldsymbol\mu{\mathbf1}{\mathbf1}}^{}\ \hat b_{{\boldsymbol\nu}}^\dagger \hat b_{\boldsymbol\mu}^\dagger \hat b_{{\mathbf1}} \hat b_{{\mathbf1}}   + U_{{\mathbf1}{\mathbf1}{\boldsymbol\nu}\boldsymbol\mu}^{}\ \hat b_{{\mathbf1}}^\dagger \hat b_{{\mathbf1}}^\dagger \hat b_{{\boldsymbol\nu}} \hat b_{\boldsymbol\mu}+\right.\\ 
 &\hspace{1 cm} \left. + V_{{\boldsymbol\nu} {\mathbf1}\boldsymbol\mu {\mathbf1}}^{}\ \hat b_{{\boldsymbol\nu}}^\dagger \hat b_{{\mathbf1}} \hat f_{\boldsymbol\mu}^\dagger \hat f_{{\mathbf1}}   + V_{{\mathbf1}{\boldsymbol\nu}{\mathbf1}\boldsymbol\mu}^{}\ \hat b_{{\mathbf1}}^\dagger \hat b_{{\boldsymbol\nu}} \hat f_{{\mathbf1}}^\dagger \hat f_{\boldsymbol\mu}\right].
\end{split}\label{eq:ContributionsDouble}
\end{equation}
\end{enumerate}
Only local contributions are taken into account and thus the spatial index $\mathbf j$ is ommited for the moment. The further intraband contributions are defined as
\begin{enumerate}
\setcounter{enumi}{3}
 \item the band energy 
\begin{equation}
 \Delta^x_{\boldsymbol\nu} =\int {\rm d}^3r\, \bar w^x_{{\boldsymbol\nu}}(\mathbf r)\left[-\frac{\hbar^2}{2m_x}\Delta+V^x(\mathbf r) \right]w_{{\boldsymbol\nu}}^x(\mathbf r),
\end{equation}
\item the intraband nearest-neighbor hopping for bosons
\begin{equation}
 J_{{\boldsymbol\nu}} =\int {\rm d}^3r\, \bar w^b_{{\boldsymbol\nu}}(\mathbf r-\mathbf{\hat e})\left[-\frac{\hbar^2}{2m_b}\Delta+V^b(\mathbf r) \right]w_{{\boldsymbol\nu}}^b(\mathbf r )
\end{equation}
and correspondingly for the fermions ${\widetilde J}_{\boldsymbol\nu}$.
\end{enumerate}

\section{Full effective first-band BFHM}\label{sec:AppendixFullHamiltonian}
Under the assumptions made in the main text (i. e., only local contributions, nearest-neighbour hopping, etc.), the final form of the effective Hamiltonian is found from equations \eqref{eq:SMatrixMultibandZwo} together with the interband couplings from \eqref{eq:ContributionSingle} to \eqref{eq:ContributionsDouble}in Markov approximation. This yields

\begin{equation}
 \begin{split}
 \hat H_{{\mathbf1}}^{\rm eff} = \hat H_{{\mathbf1}}&+\sum_{{\boldsymbol\nu}\not={\mathbf1}}\sum_{\mathbf j\mathbf d}\Biggl\{ \frac{ \left(V_{{\boldsymbol\nu}{\mathbf1}{\boldsymbol\nu}{\mathbf1}}^{}\right)^2 \mathcal I^{\mathbf d}_{bf,{\boldsymbol\nu}{\boldsymbol\nu}}}4\ { \hat b_{\mathbf j+\mathbf d}^\dagger \hat f_{\mathbf j+\mathbf d}^\dagger}{ \hat f_{\mathbf j} \hat b_{\mathbf j}}\\
&\hspace{1cm} + \left(U_{{\boldsymbol\nu}{\mathbf1}{\mathbf1}{\mathbf1}}^{}\right)^2\mathcal I^{\mathbf d}_{b,{\boldsymbol\nu}}\  {\hat b_{\mathbf j+\mathbf d}^\dagger\hat n_{\mathbf j+\mathbf d}\ }{  \hat n_{\mathbf j}\hat b_{\mathbf j} }\\
 &\hspace{1cm}  + \frac{U_{{\boldsymbol\nu}{\mathbf1}{\mathbf1}{\mathbf1}}^{}V_{{\boldsymbol\nu}{\mathbf1}{\mathbf1}{\mathbf1}}^{}\mathcal I^{\mathbf d}_{b,{\boldsymbol\nu}}  }2\   { \hat m_{\mathbf j+\mathbf d}\  \hat b_{\mathbf j+\mathbf d}^\dagger}{   \hat n_{\mathbf j}\hat b_{\mathbf j}  }\\
&\hspace{1cm} + \frac{V_{{\boldsymbol\nu}{\mathbf1}{\mathbf1}{\mathbf1}}^{}U_{{\boldsymbol\nu}{\mathbf1}{\mathbf1}{\mathbf1}}^{}\mathcal I^{\mathbf d}_{b,{\boldsymbol\nu}} }2\    {\hat b_{\mathbf j+\mathbf d}^\dagger \hat n_{\mathbf j+\mathbf d}\ } { \hat m_{\mathbf j}\hat b_{\mathbf j} }\\
&\hspace{1cm} + \frac{\left(V_{{\boldsymbol\nu}{\mathbf1}{\mathbf1}{\mathbf1}}^{}\right)^2 \mathcal I^{\mathbf d}_{b,{\boldsymbol\nu}}}4 \  {\hat m_{\mathbf j+\mathbf d}\  \hat b_{\mathbf j+\mathbf d}^\dagger}{ \hat m_{\mathbf j}\hat b_{\mathbf j}  }\\
&\hspace{1cm}+ \frac{\left(V_{{\mathbf1}{\mathbf1}{\boldsymbol\nu}{\mathbf1}}^{}\right)^2 \mathcal I^{\mathbf d}_{f,{\boldsymbol\nu}}}4\  {\hat n_{\mathbf j+\mathbf d}\ \hat f_{\mathbf j+\mathbf d}^\dagger} { \hat n_{\mathbf j} \hat f_{\mathbf j} }\Biggr\}\\
  &+ \sum_{{\boldsymbol\nu},\boldsymbol\mu\not={\mathbf1}}\sum_{\mathbf j\mathbf d}\Biggl\{ \frac{ \left(U_{{\boldsymbol\nu}\boldsymbol\mu{\mathbf1}{\mathbf1}}^{}\right)^2}4\   \mathcal I^{\mathbf d}_{bb,{\boldsymbol\nu}\boldsymbol\mu} { \left({ \hat b_{\mathbf j+\mathbf d}^\dagger} \right)^2}{\hat b_{\mathbf j}^2} \\
&\hspace{1cm}+\frac{ \left(V_{{\boldsymbol\nu}{\mathbf1}\boldsymbol\mu {\mathbf1}}^{}\right)^2}4\   \mathcal I^{\mathbf d}_{bf,{\boldsymbol\nu}\boldsymbol\mu} 
{ { \hat b_{\mathbf j+\mathbf d}^\dagger\hat f_{\mathbf j+\mathbf d}^\dagger}}  {\hat b_{\mathbf j}\hat f_{\mathbf j}}     \Biggr\}.
 \end{split}\label{eq:BthBandEffectieHamiltonian}
\end{equation}
In the Hamiltonian, the time integrals over the bosonic and fermionic correlators are defined as

\begin{align}
  \mathcal I_{b,{\boldsymbol\nu}}^{\mathbf d} &= - \frac{i}{\hbar}\int_0^\infty {\rm d}T\ 
\langle  \hat b_{{\boldsymbol\nu},\mathbf j+\mathbf d}(\tau+T)   \hat b_{{\boldsymbol\nu},\mathbf j}^\dagger(\tau)\rangle
,\\
  \mathcal I_{bf,{\boldsymbol\nu}\boldsymbol\mu}^{\mathbf d} &= - \frac{i}{\hbar}\int_0^\infty {\rm d}T\ 
\langle  \hat b_{{\boldsymbol\nu},\mathbf j+\mathbf d}(\tau+T)   \hat b_{{\boldsymbol\nu},\mathbf j}^\dagger(\tau)\rangle\times\nonumber\\
&\hspace{2.5cm}\times\langle \hat f_{\boldsymbol\mu,\mathbf j+\mathbf d}(\tau+T)   \hat f_{\boldsymbol\mu,\mathbf j}^\dagger(\tau)\rangle
,\label{eq:TimeIntegrals}
\end{align}
and correspondingly $\mathcal I_{f,{\boldsymbol\nu}}^{\mathbf d}$ and $\mathcal I_{bb,{\boldsymbol\nu}\boldsymbol\mu}^{\mathbf d}$.
The two-point correlation functions of bosons and fermions in the ${\boldsymbol\nu}$-th band read
\begin{align}
\langle  \hat b_{{\boldsymbol\nu},\mathbf j+\mathbf d}(\tau+T)   \hat b_{{\boldsymbol\nu},\mathbf j}^\dagger(\tau)\rangle_{{\boldsymbol\nu}}&=\frac 1{L^3} \sum_{\mathbf k} e^{-2\pi i \frac{ \mathbf k\cdot\mathbf d}{L^3}}  e^{\frac i\hbar T  \epsilon_{\mathbf k}^{b,{\boldsymbol\nu}}}\\
\langle \hat f_{{\boldsymbol\nu},\mathbf j+\mathbf d}(\tau+T)   \hat f_{{\boldsymbol\nu},\mathbf j}^\dagger(\tau)\rangle_{{\boldsymbol\nu}}&=\frac 1{L^3} \sum_{\mathbf k} e^{-2\pi i \frac{\mathbf k\cdot\mathbf d}{L^3}}  e^{\frac i\hbar T  \epsilon_{\mathbf k}^{f,{\boldsymbol\nu}}}.
 \end{align}

Carrying out the time integration gives
in the thermodynamic limit, which is obtained for $L\to\infty$ by setting $\mathbf\xi=\frac {\mathbf k}L$ and changing $\frac 1{L^3}\sum_{\mathbf k}$ to $\iiint {\rm d}^3\xi$ yields:
\begin{equation} 
\mathcal I_{x,{\boldsymbol\nu}}^{\mathbf d}= \frac1{(2\pi)^3}\iiint \!\!{\rm d}^3\xi\,\, \frac{e^{-i \mathbf\xi\cdot\mathbf d}}{\epsilon^{x,{\boldsymbol\nu}}(\mathbf\xi)}\label{eq:IntegralOne},
\end{equation}
\begin{equation}
 \mathcal I_{bx,{\boldsymbol\nu}\boldsymbol\mu}^{\mathbf d} = \frac{1+\delta_{\boldsymbol\nu\boldsymbol\mu}\delta_{bx}}{(2\pi)^6}\idotsint\!\! {\rm d}^3\xi\ {\rm d}^3\xi^\prime \ 
 \frac{e^{-i \mathbf\xi\cdot\mathbf d}e^{-i\mathbf\xi^\prime\cdot\mathbf d}}{\epsilon^{b,{\boldsymbol\nu}}(\mathbf\xi)+\epsilon^{x,\boldsymbol\mu}(\mathbf\xi^\prime)}. \label{eq:IntegralTwo}
\end{equation}
Here 
\begin{align}
 \epsilon^{b,{\boldsymbol\nu}}(\mathbf\xi) &= \sum_{\alpha=x,y,z}\ 2 J_{\nu_\alpha} \cos (\xi_\alpha) + \Delta_{\nu_\alpha}^b\\
 \epsilon^{f,{\boldsymbol\nu}}(\mathbf\xi) &= \sum_{\alpha=x,y,z}\  2 \widetilde J_{\nu_\alpha} \cos (\xi_\alpha) + \Delta_{\nu_\alpha}^f
\end{align}
is the energy of a boson respectively fermion in the higher band and $x$ distinguishes between bosons ($x=b$) and fermions ($x=f$).

\section{Definition of constants in Hamiltonian \eqref{eq:EffectivFirstbandHamiltonain}}\label{sec:AppendixDefinition}
As used in Hamiltonian \eqref{eq:EffectivFirstbandHamiltonain}, the full expressions of the different parameters are:\\

\noindent Density-mediated fermionic or bosonic hopping:
 \begin{multline}
J[\hat n_{\mathbf j},\hat n_{\mathbf j+\mathbf{\hat e}},\hat m_{\mathbf j},\hat m_{\mathbf j+\mathbf{\hat e}}] = \\
J-J_{nl}^b 
\left(\hat n_{\mathbf j+\mathbf{\hat e}}+\hat n_{\mathbf j}\right) -\frac{ J_{nl}^f }{2}\ \left(\hat m_{\mathbf j+\mathbf{\hat e}}+\hat m_{\mathbf j}\right)\label{DensityMediatedBosonic}
\\
  -\sum_{{\boldsymbol\nu}\not={\mathbf1}}\mathcal I^{\mathbf{\hat e}}_{b,{\boldsymbol\nu}}\Biggl\{\left(U_{{\boldsymbol\nu}{\mathbf1}{\mathbf1}{\mathbf1}}^{}\right)^2\  \hat n_{\mathbf j+\mathbf{\hat e}}  \hat n_{\mathbf j} + 
\frac{U_{{\boldsymbol\nu}{\mathbf1}{\mathbf1}{\mathbf1}}^{} V_{{\boldsymbol\nu}{\mathbf1}{\mathbf1}{\mathbf1}}^{}  }{2}\    \hat m_{\mathbf j+\mathbf{\hat e}} \hat n_{\mathbf j}\\
+\frac{V_{{\boldsymbol\nu}{\mathbf1}{\mathbf1}{\mathbf1}}^{} U_{{\boldsymbol\nu}{\mathbf1}{\mathbf1}{\mathbf1}}^{}  }{2}\    \hat n_{\mathbf j+\mathbf{\hat e}} \hat m_{\mathbf j}+ \frac{\left(V_{{\boldsymbol\nu}{\mathbf1}{\mathbf1}{\mathbf1}}^{}\right)^2 
}{4} \   \hat m_{\mathbf j+\mathbf{\hat e}}  \hat m_{\mathbf j} \Biggr\}.
\end{multline}
\begin{multline}
\widetilde J[\hat n_{\mathbf j},\hat n_{\mathbf j+\mathbf{\hat e}}]=\widetilde J-\frac{\widetilde J_{nl}}{2}\  \left(\hat n_{\mathbf j+\mathbf{\hat e}}+\hat n_{\mathbf j}\right)\label{DensityMediatedFermionic}\\
-\sum_{{\boldsymbol\nu}\not={\mathbf1}}\frac{\left(V_{{\mathbf1}{\mathbf1}{\boldsymbol\nu}{\mathbf1}}\right)^2 \mathcal I^{\mathbf{\hat e}}_{f,{\boldsymbol\nu}}}4\  \hat n_{\mathbf j+\mathbf{\hat e}} n_{\mathbf j},
\end{multline}
pair tunneling amplitude:
\begin{align}
  J^{(2)} &= \frac{ U^{\mathbf j+\mathbf{\hat e},\mathbf j+\mathbf{\hat e},\mathbf j,\mathbf j}_{{\mathbf1}{\mathbf1}{\mathbf1}{\mathbf1}} }{2}+\sum_{{\boldsymbol\nu}\not={\mathbf1}}\frac{ \left(U_{{\boldsymbol\nu}{\boldsymbol\nu} {\mathbf1}{\mathbf1}}^{}\right)^2\mathcal 
I^{\mathbf{\hat e}}_{bb,{\boldsymbol\nu}{\boldsymbol\nu}} }{2}\label{eq:CorrelatedBosonic}\\
&\hspace{0.5cm}+\sum_{\substack{{\boldsymbol\nu},\boldsymbol\mu\not={\mathbf1}\\{\boldsymbol\nu}\not=\boldsymbol\mu}} \frac{ \left(U_{{\boldsymbol\nu}\boldsymbol\mu{\mathbf1}{\mathbf1}}^{}\right)^2\mathcal I^{\mathbf{\hat e}}_{bb,{\boldsymbol\nu}\boldsymbol\mu}}{4},\notag\\
 \widetilde J^{(2)} &=\frac{V^{\mathbf j+\mathbf{\hat e},\mathbf j,\mathbf j+\mathbf{\hat e},\mathbf j}_{{\mathbf1}{\mathbf1}{\mathbf1}{\mathbf1}} }{2} + \sum_{{\boldsymbol\nu}\not={\mathbf1}}\frac{ \left(V_{{\boldsymbol\nu} {\mathbf1}{\boldsymbol\nu} {\mathbf1}}^{}\right)^2\mathcal 
I^{\mathbf{\hat e}}_{bf,{\boldsymbol\nu}{\boldsymbol\nu}} }4\label{eq:CorrelatedFermionic}\\
&\hspace{0.5cm}+\sum_{\substack{{\boldsymbol\nu},\boldsymbol\mu\not={\mathbf1}\\{\boldsymbol\nu}\not=\boldsymbol\mu}} \frac{ \left(V_{{\boldsymbol\nu} {\mathbf1} \boldsymbol\mu{\mathbf1}}\right)^2}4\   
\mathcal I^{\mathbf{\hat e}}_{bf,{\boldsymbol\nu}\boldsymbol\mu},\notag
\end{align}
renormalized two-particle interactions:
\begin{align}
 U_2&= U+ \sum_{{\boldsymbol\nu}\not={\mathbf1}}  \left(U_{{\boldsymbol\nu}{\mathbf1}{\mathbf1}{\mathbf1}}^{}\right)^2\mathcal I^{\mathbf0}_{b,{\boldsymbol\nu}}+\sum_{\boldsymbol\nu,\boldsymbol\mu\not={\mathbf1}} \frac14{ \left(U_{{\boldsymbol\nu}\boldsymbol\mu{\mathbf1}{\mathbf1}}^{}\right)^2\mathcal I^{\mathbf 0}_{bb,{\boldsymbol\nu}\boldsymbol\mu}},\\
V_2 &= V + \sum_{{\boldsymbol\nu}\not={\mathbf1}} \frac{\left(V_{{\boldsymbol\nu}{\mathbf1}{\boldsymbol\nu}{\mathbf1}}^{}\right)^2 \mathcal I^{\mathbf0}_{bf,{\boldsymbol\nu}{\boldsymbol\nu}}}2\notag\\
&\hspace{0.5cm}+ \sum_{{\boldsymbol\nu}\not={\mathbf1}}\left\{\frac {\left(V_{{\boldsymbol\nu}{\mathbf1}{\mathbf1}{\mathbf1}}^{}\right)^2 \mathcal I_{b,{\boldsymbol\nu}}^{\mathbf0}}2+\frac{\left(V_{{\mathbf1}1{\boldsymbol\nu}1}^{}\right)^2\mathcal I_{f,{\boldsymbol\nu}}^{\mathbf0}}2\right\}\label{eq:RenormalizedInteraction-f}\\
&\hspace{0.5cm}+\sum_{\substack{{\boldsymbol\nu},\boldsymbol\mu\not={\mathbf1}\\{\boldsymbol\nu}\not=\boldsymbol\mu}} \frac{ \left(V_{{\boldsymbol\nu}{\mathbf1} \boldsymbol\mu{\mathbf1}}\right)^2}2\   \mathcal I^{\mathbf 0}_{bf,{\boldsymbol\nu}\boldsymbol\mu},\notag
\end{align}
three-body interactions
\begin{align}
 U_3 &= 6\sum_{{\boldsymbol\nu}\not={\mathbf1}}{\left(U_{{\boldsymbol\nu}{\mathbf1}{\mathbf1}{\mathbf1}}\right) ^2\mathcal I^{\mathbf0}_{b,{\boldsymbol\nu}}},\\  
  V_3&=  \sum_{{\boldsymbol\nu}\not={\mathbf1}}\left\{ U_{{\boldsymbol\nu}{\mathbf1}{\mathbf1}{\mathbf1}}^{} V_{{\boldsymbol\nu}{\mathbf1}{\mathbf1}{\mathbf1}}^{}\mathcal I^{\mathbf0}_{b,{\boldsymbol\nu}} +\frac{\left(V_{{\mathbf1}{\mathbf1}{\boldsymbol\nu}{\mathbf1}}^{}\right)^2\mathcal I_{f,{\boldsymbol\nu}}^{\mathbf0}}4\right\}.
  \end{align}
  
  \section{Lattice effects}\label{sec:AppendixLatticeEffects}

The lattice potentials for bosons and fermions are both created by the same laser field and the only externally 
controllable parameter is the intensity of this lattice laser. In order to see how the parameters of the effective
lattice model, such as tunneling rates and interaction constants depend on this laser intensity one needs to take
into account that there is always a fixed ratio ${\widetilde f}$ between the bosonic and fermonic
potential depths for given atomic species and transitions.   To determine ${\widetilde f}$
we note that  the optical lattice is generated by an off-resonant standing laser field. The potential itself results from the ac-Stark shift. As shown in \cite{Grimm2000}, it is given by
\begin{equation}
 V_{\rm pot}(\mathbf r) = \frac{3\pi c^2}{2} 
\left(\frac{\Gamma_{D_1}}{\omega_{0,D_1}^3\Delta_{D_1}}+\frac{2\Gamma_{D_2}}{\omega_{0,D_2}^3\Delta_{D_2}}\right) \  I(\mathbf r)\label{eq:ac-stark}
\end{equation}
in rotating wave approximation for a typical alkali D-line doublet, where each line contributes independently if the laser is sufficiently far detuned from the atomic transitions. The important parameters are the decay rates $\Gamma_{D_{1,2}}$ of the excited states, $\Delta_{D_{1,2}}=\omega_{\rm laser}-\omega_{0,D_{1,2}}$ the detunings of the laser frequency $\omega_{\rm laser}$ from the atomic transition frequencies $\omega_{0,D_{1,2}}$ and $I(\mathbf r)= I_0\ \sin^2(\mathbf k \mathbf r)$ the laser intensity.

Conveniently, all energies in the system are normalized to the recoil energy of the bosonic species given by $ E_{\rm rec}^{b} = \frac{\hbar^2 k^2}{2 m_b}$.  The wavenumber $k$ depends on the chosen optical lattice. The
(normalized) lattice potential for the bosons thus reads $V_{\rm lat}^b(\mathbf r) = \eta_b \sin^2(\mathbf k\mathbf r)$. 
It is useful to rewrite the optical lattice potential for the fermionic atoms with respect to the bosonic optical lattice as 
$V_{\rm lat}^f(\mathbf r) =  \eta_f \ \sin^2(\mathbf k\mathbf r)$, where $\eta_f ={\widetilde f}\, \eta_b$. From eq. \eqref{eq:ac-stark} we find
\begin{equation}
 \widetilde f = 
\frac{\frac{\Gamma_{D_1}^f}{\left(\omega_{0,D_1}^f\right)^3\Delta_{D_1}^f}+\frac{2\Gamma_{D_2}^f}
{\left(\omega_{0,D_2}^f\right)^3\Delta_{D_2}^f}}{\frac{\Gamma_{D_1}^b}{\left(\omega_{0,D_1}^b\right)^3
\Delta_{D_1}^b}+\frac{2\Gamma_{D_2}^b}{\left(\omega_{0,D_2}^b\right)^3\Delta_{D_2}^b}}\ .\label{eq:PrefactorLatticeDepth}
\end{equation}
At this point, we specify the experimental system.  In the previous discussions, we analyzed the experiment reported in \cite{Best2009} and use the parameters given there. A mixture of bosonic  $^{87}$Rb and fermionic $^{40}$K is cooled and put into an optical lattice with $\sigma_L=755$ nm.  For Rubidium and Potassium, the transition wavelengths and decay rates are given by
\begin{align}
 \sigma_{D_1}^{K} &= 766.5\ {\rm nm} \hspace{1cm} & \sigma_{D_1}^{Rb} &= 795.0\ {\rm nm}\notag\\
 \Gamma_{D_1}^K&= 38.7\times 10^6\   {\rm Hz} &  \Gamma_{D_1}^{Rb}&= 36.1\times 10^6\   {\rm Hz}\\
 \sigma_{D_2}^{K} &= 769.9\ {\rm nm} \hspace{1cm} & \sigma_{D_2}^{Rb} &= 780.2\ {\rm nm}\notag\\
 \Gamma_{D_2}^K&= 38.2\times 10^6\   {\rm Hz} &  \Gamma_{D_2}^{Rb}&= 38.1\times 10^6\   {\rm Hz}\notag.
\end{align}
Using these values, $\widetilde f$ in equation \eqref{eq:PrefactorLatticeDepth} evaluates to $\widetilde f = 2.04043$, 
which means, that the fermionic lattice potential, in terms of the bosonic recoil energy is twice as deep as the bosonic one.
For the calculation of the Wannier functions of bosons and fermions one has to take into account however also the
different masses of the particles. Expressing the  Schr\"odinger equation for the single-particle fermionic
wavefunction $\Phi_f(\mathbf r)$ in terms of the bosonic quantities $\eta_b$ and $m_b$, one finds
\begin{equation}
\left[ -\frac{\hbar^2}{2 m_b}\Delta + \frac{m_f}{m_b}\widetilde f\ \eta_b\ \sin^2(\mathbf k\mathbf r) \right]\ \Phi_f(\mathbf r) =  \frac{m_f}{m_b}E\ \Phi_f(\mathbf r).\label{eq:SchroedingerEquationFermionsToBosons}
\end{equation}
One recognizes that the difference between the fermionic Wannier functions and the bosonic ones is determined
only by the factor  $\frac{m_f}{m_b}\widetilde f$. Since for the experiment in \cite{Best2009}
\begin{equation}
 \frac{E^f_{rec}}{E^b_{rec}} = \frac{m_b}{m_f} = 2.175
\end{equation}
the factor $\widetilde f$ is almost compensated,  $\frac{m_f}{m_b}\widetilde f=0.93$. Thus the bosonic and fermionic Wannier functions are to a good approximation identical with a maximal overlap. Nevertheless, figure \ref{fig:TransitionShiftScatteringLengthBands} also display results including a mismatch of the bosonic and fermionic Wannier functions, depicted by the gray shaded regions. The upper (lower) boundary on the attractive side and the lower (upper) boundary on the repulsive site corresponds to the results for a mismatch of $\frac{m_f}{m_b}\widetilde f=0.7$ $(\frac{m_f}{m_b}\widetilde f=1.3$), indicating the importance of a good control of the mismatch in the precise determination of the transition shift.\\

\end{appendix}


\bibliographystyle{apsrev}
\bibliography{BibTex_Alexander_Mering}

\end{document}